# Mini-review: Probing the limits of extremophilic life in extraterrestrial environment-simulated experiments

Claudia LAGE[1]*, Gabriel DALMASO[1], Lia TEIXEIRA[1], Amanda BENDIA[1], Ivan PAULINO-LIMA[2], Douglas GALANTE[3], Eduardo JANOT-PACHECO[3], Ximena ABREVAYA[4], Armando AZÚA-BUSTOS[5], Vivian PELIZZARI[6], Alexandre ROSADO[7]

[1] Laboratório de Radiações em Biologia, Instituto de Biofísica Carlos Chagas Filho, Universidade Federal do Rio de Janeiro, Brazil

[2] NASA-Ames Research Center, USA

[3] Instituto de Astronomia, Geofísica e Ciências Atmosféricas, Universidade de São Paulo, Brazil

[4] Instituto de Astronomía y Física del Espacio, Universidad de Buenos Aires - CONICET, Argentina

[5] Pontificia Universidad Catolica de Chile, Chile

[6] Instituto Oceanográfico, Universidade de São Paulo, Brazil

[7] Instituto de Microbiologia Prof Paulo Góes, Universidade Federal do Rio de Janeiro, Brazil

* corresponding author
Instituto de Biofisica Carlos Chagas Filho - UFRJ
Centro de Ciencias da Saude, Bldg G
Av. Carlos Chagas Filho, 373
Cidade Universitaria
21941-902 Rio de Janeiro RJ BRAZIL
Phone numbers: +55 21 2562 6576 (office)
        9964 6866 (mobile)
        2280 8193 (FAX)
E-mail:    lage@biof.ufrj.br

**Abstract**

  Astrobiology is a brand new area of science that seeks to understand the origin and dynamics of life in the universe. Several hypotheses to explain life in the cosmic context have been developed throughout human history, but only now technology has allowed many of them to be tested. Laboratory experiments have been able to show how chemical elements essential to life, carbon, nitrogen, oxygen and hydrogen combine in biologically important compounds. Interestingly, these compounds are found universally. As these compounds were combined to the point of originating cells and complex organisms is still a challenge to be unveiled by science. However, our 4.5 billion years- old solar system was born within a 10- billion years- old universe. Thus, simple cells like microorganisms may have had time to form in planets older than ours or other suitable molecular places in the universe. One hypothesis to explain the origin of life on Earth is called panspermia, which predicts that microbial life could have been formed in the universe billions of years ago, traveling between planets, and inseminating units of life that could have become more complex in habitable planets like ours. A project designed to test viability of extremophile microorganisms exposed to simulated extraterrestrial environments is ongoing at the Carlos Chagas Filho Institute of Biophysics to test whether microbial life could withstand those inhospitable environments. Ultra- resistant (known or novel ones) microorganisms collected from terrestrial extreme environments, extremophiles, have been exposed to intense radiation sources simulating solar radiation (synchrotron accelerators), capable of emitting in a few hours radiation equivalent of million years accumulated doses. The results obtained in these experiments have revealed a remarkable resistance of extremophilic bacteria and archaea against different radiation sources (VUV, solar wind simulants, X rays) whenever protected by microsized carbonaceus grains. Altogether, the collected data suggest the interesting possibility of the existence of microbial life beyond Earth and its transfer among habitable bodies, which we have called *microlithopanspermia*.

**Keywords:** panspermia, extremophiles, UV, cosmic dust

**The panspermia hypothesizing one origin of life on Earth**

The panspermia hypothesis (Arrhenius, 1903) ("seeds everywhere") suggests that life could be spread everywhere. The panspermia hypothesis assumes that spreading life elsewhere depends on three steps: (i) the escape step, *i.e.*, ejection of contaminated planetary material towards space, normally caused by a large impact on the parent planet; (ii) the journey in space through time scales comparable to those experienced by the Martian meteorites (estimated as 1–15 million years); and (iii) the landing process in a manner to afford non-destructive deposition of the biological material on a recipient planet (Horneck *et al.*, 2003). In the last decade many experiments have been used to evaluate the microorganism survival in space conditions and also different species of microorganisms have been utilized for that (Horneck *et al.*, 2010; Olsson-Francis & Cockell, 2010).

The first step of panspermia hypothesis have already been tested by exposing spores of *B. subtilis*, cells of *Chroococcidiopsis* and thalli and ascocarps of the lichen *Xanthoria elegans* to strong shock pressures ranging from 5 to 40 GPascals. Their results support the hypothesis that biological material could be successfully ejected from planets in a way that seeding of early Earth might have ocurred (Horneck *et al.*, 2008).

The second step, one of the many possible mechanisms for panspermia, is called lithopanspermia. It was initially suggested as a mechanism where meteorites passing by Earth's low-orbit in a shallow angle allows them to collect viable microorganisms present in the upper atmosphere (Nicholson, 2009), and meteorites would function as vehicles transferring life forms throughout space. Calculations by Mileikowsky *et al.* (2000) have predicted that any rock-captured microorganisms such as *D. radiodurans* and *Bacillus sp.* should be shielded against space radiation inside rocks of the order of at least 0.33m size, to keep a viable minimal population during time ranges suitable to afford Mars–Earth interplanetary travel (about 1 million years). The main constraint concerning lithopaspermia is argued on the basis of secondary radiation effects caused by charged particles which enhance energy absorption by a massive rock fragment. The step of long-term survival in space awaits more concrete results to reinforce the panspermia hypothesis.

The third and decisive step, re-entry on a habitable body, was investigated by Cockell *et al.* (2007), who exposed an endolithic photosynthetic organism *Chroococcidiopsis sp.* in the ESA's STONE experiment. Inoculated into a gneissic rock sample, this cyanobacterial sample did not resist the speedy re-entrance into the Earth atmosphere since extreme heating reached 5 mm deep in the rock.

**The task of surviving radiation under simulation experiments**

Extremophilic organisms can survive and proliferate in non-mesophilic conditions such as extreme temperatures, salinity, pressure, pH, desiccation and radiation (Rothschild & Mancinelli, 2001), been useful to a wide range of research areas, from basic microbiology to the development of industrial technology. The poly-extremophilic bacteria *Deinococcus radiodurans* is one of the most radiation resistant organisms yet discovered, capable of surviving acute doses of ionizing radiation (IR) exceeding 15kGy (Daly *et al.*, 1994) and growing under chronically-delivered gamma radiation (60Gy·h$^{-1}$) (Daly, 2000). The molecular mechanisms underlying extreme radiation resistance in *D. radiodurans* have been the subject of basic research for decades (Cox & Battista, 2005; Makarova *et al.*, 2001), now it is conceived that its resistance arises from the summing of anti-oxidant activities (Blasius *et al.*, 2008).

The resistance of *D. radiodurans* also applies to ultraviolet radiation, specifically to shortwave ultraviolet (UV) radiation, with the dose leaving ~1/3 survivors (LD$_{37}$) approaching 600J·m$^{-2}$ (Battista, 1997). Nevertheless, its extremely efficient repair of critically lethal double strand breaks (DSB) resealing millions DSB (Zahradka *et al.*, 2006) is assumed to be in the basis of its radioresistance. Other anti-oxidant functions, if overcome by excessive oxidation, render cells much more sensitive to damage (Daly *et al.*, 2007). If cellular post-IR recovery rely on the activity of rescue proteins, one mechanism described to prevent IR-induced protein oxidation is the accumulation of Mn$^{+2}$ complexes in resistant organisms, decreasing intracellular concentration of damaging reactive oxygen species (ROS) formed during irradiation (Daly, 2009).

Extremophile microorganisms have been largely tested to experimentally measure the limits for life to exist in order to assist the search for life in the Universe. The Solar System has a number of astrobiological interesting locations - ranging from planet Mars, the moons Europa, Titan and Enceladus, to comets and asteroids, which may be analogous to large compartments carrying living matter in a state of suspended animation for millions years, just waiting for landing on a habitable location to sprawl.

**Brazilian experimental setup to test a novel concept in the panspermia scenario**

Current lithopanspermia models admit the possibility of interplanetary transport of endolithic microbes between donor and recipient telluric planets (Nicholson *et al.*, 2005). The planned experimental setup designed by our team envisaged as a first hypothesis the "n>1" perspective, *i.e.*, such process could have occurred between habitable planets in our own solar system (including Mars). In other words, we conceived testing the resistance of an extremely radiation resistant microbe exposed to as many scenarios as possibly existing during a putative Mars-Earth interplanetary travel to which any living cell could have to cope with.

Our group´s goal was set on performing simulation experiments concerning the critical step of microbial resistance to long-term exposure to extraterrestrial simulated conditions. A simulated source of solar radiation present in interplanetary space was found in the Brazilian Synchrotron Light Laboratory (LNLS) Toroidal Grating Monochromator (TGM) beamline (Cavasso Filho *et al.*, 2007). The beam was focused to coincident samples inside a vacuum chamber, receiving an energy range from 0.1 to 21.6eV (photons flux of ~10$^{14}$ s$^{-1}$) with a spectral region comprising from infrared (IR) to vacuum-ultraviolet (VUV). The non-spore forming radioresistant bacteria *Deinococcus radiodurans* was prepared as a deeply dehydrated cell powder and exposed to increasing doses of VUV synchrotron radiation mimicking the solar spectrum. If trapped into a porous carbon tape (Figure 1) and shielded by cell multi-layers surrounded by organic matrix material, *D. radiodurans* was shown to survive doses equivalent to ~420 days exposure to unfiltered VUV at 1 AU (Paulino-Lima *et al.*, 2010b). Under an astrobiological perspective, this data can be interpreted as one possible context for incoming microparticulate, extremophile-contaminated material to disperse on top atmospheric layers, with cells remaining viable for more than one year suspended before landing on the planet surface at any point of Earth history.

In another experimental setup, cells of *D. radiodurans* were exposed to different sources of simulated charged particles found in solar wind. Naked cells or cells mixed with dust grains (basalt or sandstone) differing in elemental composition were exposed to electrons, protons, and carbon ions aiming at determine the probability of cell survival under solar wind particles bombardment. The results of this study indicate that low-energy particle radiation (2–4keV), typically present in the slow component of solar wind, had no effect on dehydrated cells, even if exposed to equivalent energies accumulated if exposed after some 1000 years at 1 AU. Higher energy carbon ions (200keV) found in solar flares would inactivate 90% of exposed cells. These results show that, compared to the highly deleterious effects of VUV radiation, solar wind charged particles are relatively less damaging, and organisms protected by dust grains from UV radiation would also be protected from simulated bombardment by charged particles (Paulino-Lima *et al.*, 2011).

Following the study on radio-resistant microorganisms in the context of panspermia, the salt-resistant haloarchaea *Natrialba magadii* and *Haloferax volcanii* were exposed to the same VUV TGM beamline, dehydrated and under vacuum. *N. magadii* was remarkably resistant to high vacuum with a survival fraction larger than the one of *D. radiodurans*, differently from the survival observed for *H. volcanii*, which was much more sensitive. Radiation resistance profiles were similar for both haloarchaea and *D. radiodurans* for VUV doses up to 150J·m$^{-2}$. For doses higher than 150J·m$^{-2}$ there was a significant drop in survival of both haloarchaea, and in particular, no *H. volcanii* growth was detected after any further dose increment. Survival for *D. radiodurans* was 1% after exposure to the higher VUV dose (1.35kJ·m$^{-2}$) while *N. magadii* survival approached some 0.1%, yet a remarkable score for this organism. Such survival fractions are discussed regarding the possibility of interplanetary transfer of viable salt-loving microorganisms from particular extraterrestrial salty environments such as the planet Mars and the Jupiter's moon Europa to Earth. This is the first work reporting survival of haloarchaea under simulated interplanetary conditions (Abrevaya *et al.*, 2011).

In order to better provide shielding for microorganisms, micro particles or even dust particles should be more effective to transfer living matter in interplanetary space. The hypothesis of micro/dust particles also should provide a minimal impact upon atmosphere reentry, allowing the microorganism to reach the planetary surface.

**Peering Earth extreme environments in search of novel extremophiles**

The Antarctic region is usually remembered by its extreme environmental conditions like low temperatures, low nutrient availability, high UV radiation flux, freeze and thawing cycles and long Sun-shaded periods. The Antarctic environment offers unique opportunities for research concerning microbial diversity and it is one of the untouched, least characterized environments available to research. Its many ecologically interesting environments include permanently frozen soil (permafrost), mineral soils, frozen lakes, snow and glacial ice. Microorganisms sampled from these peculiar fields can help unveiling how broad can be the limits for life to exist, in a particularly interesting extraterrestrial analog environment.

Antarctic bacteria have been used as model organisms in the study of extremophilic biology. One particular goal that has been followed by our group is the screening for UV resistance among Antarctic microorganisms. Extraordinarily, these organisms are able to survive to both ionizing and non-ionizing (UV) radiation-rich environments, which could otherwise be lethal to others. UV radiation resistance has thus been used to select microorganisms to be subjected to extraterrestrial conditions in future simulation experiments.

In a new recent project developed by our team, novel bacterial isolates from Antarctic soils were analyzed. Besides the UV screening, these bacteria were also characterized by their ability to produce biotechnologically interesting enzymes with potential industrial uses. This approach was set in a view to streamline astrobiology and biotechnology for more robust bioprocesses.

The other peered astrobiologically interesting environment was the Atacama desert in Chile, considered similar to Mars due to parameters such as extremely low humidity and intense solar UV radiation (Connon *et al.*, 2007; Navarro-Gonzalez *et al.*, 2003) Its location comprises part hyper-arid desert of Peru, Chile, and the Chilean portion extends between 1°N and 37°S, incorporating the arc Peru, northern Chile and parts of western Andes cordillera (Hartley *et al.*, 2005). The nearly absent clouds contribute to the occurrence of an yearly intense flux of solar radiation over the entire Atacama. Altogether, in the Atacama environment are cast many Martian features. In the case of Mars, high levels of poorly filtered UV radiation reaches the surface (Schuerger *et al.*, 2003), contrary to what occurs on Earth, where wavelengths below 300nm are atmospherically attenuated. As such, any putative microorganism present on the Martian surface is subject to constant exposure to high fluxes of UV light highly deleterious (Cockell, 2001).

In a parallel set of experiments, procedures were developed for isolation of UV resistant microorganisms of soil from the Atacama Desert. As for the Antarctic project, the goal of this project was to isolate novel UV resistant microorganisms resistant to subject them to future experimental simulations. Among some 30 richly pigmented Atacama isolates, a particular one was shown to be highly UV resistant, although not to the same extent as *D. radiodurans* (Paulino-Lima *et al.*, submitted).

The screening for novel extremophile species has been incorporated to the group as a task to investigate strategies to survive to extraterrestrial-like environments. In the near future, a new chamber under construction in the Astrobiology Laboratory in São Paulo-Brazil will allow deeper qualitative investigations on such mechanisms, besides raising cues for future advising on interplanetary protection procedures (Paulino-Lima, 2010a).

### *Deinococcus radiodurans* surprises again

In another branch of applied research, *D. radiodurans* was shown to be a prospective candidate for bioremediation of radioactive environmental waste sites, based on its ability to grow and functionally express cloned genes during exposure to chronic gamma irradiation (Brim *et al.*, 2000; Brim *et al.*, 2006). For example, a clone of *D. radiodurans* was engineered to completely degrade aromatic hydrocarbons (e.g., toluene) (Lange *et al.*, 1998) and reduce toxic metals (e.g., $Cr^{6+}$, $Hg^{2+}$) in the presence of $^{137}Cs$ (Brim *et al.*, 2006).

Under an astrobiological perspective, there are hydrocarbon-rich reservoirs in other bodies of the solar system. Titan, Saturn's largest natural satellite, is one of them, with remarkable microbial-habitable properties (Niemann *et al.*, 2005). The presence of methane and other hydrocarbons in its dense atmosphere (Atreyaa *et al.*, 2006) attenuates surface reaching UV radiation. Upper atmospheric UV causes those reactive compounds to precipitate over the surface creating lakes and rivers of hydrocarbon byproducts (Lorenz *et al.*, 2008; Mitri *et al.*, 2007; Perron *et al.*, 2006; Stofan *et al.*, 2007).

In this field of search we are currently investigating if intrinsically biological properties of the poly-extremophilic bacterium *D. radiodurans* could enable it to survive (and even metabolize) complex organic mixtures, avoiding the need for genetic engineering techniques. Microbial growth was observed after incubation with complex hydrocarbon mixtures, and *D. radiodurans* was seen to tolerate them (Dalmaso *et al.*, 2011).The results obtained in this study demonstrate the proliferation ability of *D. radiodurans* in an extreme carbon-poor environment, with high concentration of organic solvents. Moreover, these results suggest the potential use of a new class of microorganisms in the biodegradation and bioremediation of contaminated areas.

Aside from the interesting perspective for application of a naturally resistant microorganism in bioremediation procedures, our initial positive results also suggest that an extremophilic microorganisms such as *D. radiodurans* could proliferate in hydrocarbon-rich sites, like Titan, with implications for astrobiological research. Such outcomes support the notion that carbon-based life might dwell on alternative solvent ponds elsewhere.

**The group contribution in the existing background on experimental simulations in Astrobiology**

The idea that meteoritic material could thrive viable micro-organisms through long interplanetary journeys was conceived possible under many particular aspects. From theoretical Monte Carlo simulations by Mileikovsky *et al.* (2000) predicting that meter-sized meteoritic mineral rocks could really afford radiation shielding to biologically active material in a Mars-Earth transfer up to the broader supposition of microbial contaminated dust spreading out across the Galaxy (overviewed in Parsons, 1996). Many issues still hang on the ability of microbial species to survive extraterrestrial conditions. Findings from the G. Horneck group have added experimental input on this subject, stressing that viable bacterial spores should count on unfiltered radiation shielding by superficial dead cells layers (EXOBIOLOGIE experiment onboard the MIR station; Rettberg *et al.*, 2002). Extra cellular covertures, provided by inorganic dry medium and/or mineral material were also shown to enhance protection, with the remark that such chemical shielding should be effective only against UV radiation (Horneck *et al.*, 2001).

As an advance on this issue, our group took into account another piece of evidence observed in micrometer-size meteorites: many of them do not seem to undergo overheating ablation depending on their infall conditions on an atmospheric-harboring planet (Coulson, 2006). The high survival rates scored after challenging a highly extremophilic non-spore forming species with either VUV photons or solar-wind-like particles shielded by simulated micrometeoritic material led the panspermia hypothesis a step forward. Altogether, it is the contribution of this Brazilian group to have put together the determination that micrometeoritic material could shield viable extremophilic micro-organisms for long-time stand in space conditions at the same time skipping overheated re-entry into atmospheric planets.

**Concluding remarks**

Previous experiments done by Dose *et al.* (1995) had already revealed that spores, fungal conidia or vegetative cells (*D. radiodurans*) exposed to space conditions (vacuum and unshielded solar radiation) as cells monolayers were shown to be extremely sensitive; on the other hand, biological material assembled in multilayers or macroscopic clusters can resist for months or years, even if exposed to full solar light. Top cell layers are supposed to become inactivated, while keeping cached cells rescued from UV damage and also partially from dehydration. Osman *et al.* (2008) observed significant survival of spore-forming bacteria after irradiation with full spectrum Martian UV irradiation if shielded by soil particles (<60 mm) from the Atacama Desert. In addition, Pogoda de la Vega *et al.* (2007) have found that even nano-particles can afford survival of *D. radiodurans* upon UV irradiation under simulated Martian conditions.

The main challenge concerning lithopaspermia are the bystander effects due to absorption of high-energy charged (HZE) particles, showering secondary electrons, shock waves and additional thermophysical events (Paulino-Lima *et al.*, 2010a). Large rock fragments indeed can amount higher accumulated doses during a long-term stand in space. Interestingly, micro particles could hypothetically afford a better shielding for microorganisms in the transfer of living matter across interplanetary space. Therefore, hypothetical microbes within particles would be mostly inactivated by direct hits; yet little inactivation is expected to occur because of the low radiation flux (Horneck *et al.*, 2010). The interplanetary travel on hypothetic micro/dust particles should also be advantageous because of the minimal shock impact upon atmosphere re-entry, allowing nomad microorganisms to reach the planetary surface. Experimental evidence indicates that small particles containing bacteria could survive the temperature regimens imposed during re-entry into Earth's lower atmosphere (Coulson, 2004).

The collection of results obtained by our team (summarized in Table 1) has joined the concepts of cells multilayer shielding rescued inside dust microparticulate material, similarly to existing carbon or silicate cosmic dust. Such microenvironment seems to escape radiation direct hits while in long-term standing in space, as well as it does not appear to suffer from re-entry ablation on top of an atmosphere-containing recipient planet. If, after all gathered evidence, the panspermia hypothesis does not hold valid due to the really harsh extraterrestrial environment, at least our dusty Earth-borne contaminated particles may be widespread in our Galaxy, inseminating other putatively flourishing bodies with extremophilic life.

**Acknowledgements:** Coordenação de Aperfeiçoamento de Pessoal de Nível Superior (CAPES), Conselho Nacional de Pesquisas e Desenvolvimento Tecnológico (CNPq), Fundação Carlos Chagas Filho de Amparo a Pesquisa do Estado do Rio de Janeiro (FAPERJ) and Laboratório Nacional de Luz Síncrotron (LNLS-CNPEM) for research Grants and fellowships. This project was also funded by the Ministry of Science, Technology & Innovation and CNPq joint program of National Institutes of Science and Technology (INCT), INEspaço.

Table 1. Summary of results observed under extraterrestrial-simulated environments

| Organism | Experimental Simulation | Outcome | Reference |
|---|---|---|---|
| *D. radiodurans* | dehydration, vacuum, VUV, carbon tape shielding | 1% survival after 12kJ·m$^{-2}$ VUV | (Paulino-Lima *et al.*, 2010b) |
| *D. radiodurans* | dehydration, vacuum, VUV, unshielded | 1% survival after 1.3kJ·m$^{-2}$ VUV | (Abrevaya *et al.*, 2011) |
| *D. radiodurans* | dehydration, vacuum, solar wind charged particles, dust powder shielding | 10% survival after 1kGy of 200keV protons | (Paulino-Lima *et al.*, 2011) |
| *D. radiodurans* | saline culture medium added of organic hydrocarbons (solvents) | 1% survival after 45 days exposure | (Dalmaso *et al.*, 2011) |
| *Chroococcidiopsis sp.* | dehydration, vacuum, solar wind charged particles, shielding | no survival after 1kGy of 200keV protons | (Paulino-Lima *et al.*, 2010a) |
| *N. magadii* | dehydration, vacuum, VUV, unshielded | 0.1% survival after 1.3kJ·m$^{-2}$ VUV | (Abrevaya *et al.*, 2011) |
| *H. volcanii* | dehydration, vacuum, VUV, unshielded | No survival after 150J·m$^{-2}$ VUV | (Abrevaya *et al.*, 2011) |
| Atacama Desert isolate | shortwave UV | 0.1% survival after 500J·m$^{-2}$ UV | (Paulino-Lima *et al.*, submitted) |
| Antarctica isolate | shortwave UV | 1% survival after 600J·m$^{-2}$ UV | (Bendia *et al.*, unpublished results) |

**Cited bibliography:**